\documentclass[twocolumn,superscriptaddress,floatfix]{revtex4}
\usepackage{amsmath,amssymb,bm}
\usepackage{graphicx}
\usepackage{float,braket}
\usepackage{color,dsfont}
\usepackage{appendix}

\begin{document}

\title{Practical long-distance quantum key distribution\\ through concatenated entanglement swapping\\
with parametric down-conversion sources}

\author{Aeysha Khalique}
\affiliation{%
	School of Natural Sciences, National University of Sciences and Technology,
	H-12 Islamabad, Pakistan%
	}
\affiliation{%
	Hefei National Laboratory for Physical Sciences at Microscale and Department of Modern Physics, University of Science and Technology of China, Hefei, Anhui 230026, China%
	}
\affiliation{%
	Shanghai Branch, CAS Center for Excellence and Synergetic Innovation Center in Quantum Information and Quantum Physics, University of Science and Technology of China, Shanghai 201315, China%
	}
\author{Barry C. Sanders}
\affiliation{%
	Hefei National Laboratory for Physical Sciences at Microscale and Department of Modern Physics, University of Science and Technology of China, Hefei, Anhui 230026, China}
\affiliation{%
Shanghai Branch, CAS Center for Excellence and Synergetic Innovation Center in Quantum Information and Quantum Physics, University of Science and Technology of China, Shanghai 201315, China%
	}
\affiliation{%
	Institute for Quantum Science and Technology, 
	University of Calgary, Alberta T2N 1N4, Canada}
\affiliation{%
	Program in Quantum Information Science,
	Canadian Institute for Advanced Research, Toronto, Ontario M5G 1Z8, Canada
	}

\date{\today}

\begin{abstract}
We develop a theory for long-distance quantum key distribution based on concatenated entanglement swapping using parametric down-conversion sources and show numerical results of our model.
The model incorporates practical resources including multi-pair sources,
inefficient detectors with dark counts and lossy channels.
We calculate the maximum secret key-generation rate
for up to three entanglement swapping stations
by optimizing over resource parameters,
and our numerical simulation shows that the range of quantum key distribution can in principle
be markedly increased but at the expense of an atrociously unfeasible secret key-generation rate;
however,
the upper bound of our key rates closely approach the Takeoka-Guha-Wilde upper bound.
Our analysis demonstrates the need for new technology such as quantum memory to synchronize
photons and our methods should serve as a valuable component for accurately modelling
quantum-memory-based long-distance quantum key distribution. 
\end{abstract}

\maketitle

\section{Introduction}
\label{sec:intro}

Secure communication over public channels is vital for enabling effectively private information transfer.
Two steps are involved: establishing a key
and then using the key for encryption and decryption~\cite{Singh99}.
The security of existing approaches to constructing a shared key between two parties relies on the
computational complexity of the secret-key-generation algorithm,
but this security would be compromised if fast decryption algorithm were discovered
(unlikely but not provably forbidden) or if a scalable quantum computer were available.
Quantum key distribution (QKD) aims to provide information-theoretic security,
in other words to create keys that are safe against computational attacks~\cite{GRT+02}.

One key challenge for QKD is to reach long distances,
which practically means extending beyond hundreds of kilometres
with a good secret key-generation rate (SKGR)
(better than kilobits per second) and low error rates,
with less than 11\% to be corrected by one way classical communication~\cite{SP00}.
Two distinct approaches to QKD exist:
one protocol relies on one party sending qubits to the other,
exemplified by the 1984 Bennett-Brassard (BB84) protocol~\cite{BB84},
and the other based on the two parties sharing and reconciling an entangled state
exemplified by the 1991 Ekert (E91)~{\cite{Ekert91}
and 1992 Bennett-Brassard-Mermin (BBM92)~\cite{BBM92} protocols.

The BBM92 protocol is particularly amenable to the quantum relay method~\cite{CGR05},
which relies on entanglement swapping (ES)~\cite{PBW+98} to extend the distance
QKD can reach~\cite{GRT+02}.
Our theory treats practical QKD,
where the term ``practical'' refers to the theory accommodating
the reality of photon sources and detectors including multi-photon events, dark counts and losses~\cite{BLMS00, MQZ+05}.

Here we study the reach of practical quantum-relay-based
long-distance QKD using a theory that treats photon sources, communication channels and detectors realistically.
Our model shows that QKD could reach over $800$~km in principle but at atrociously bad SKGR.
An important result of our theory is the underscoring
of the need for quantum memory~\cite{LST09, SAA+10}
to synchronize photons so that the poor rates obtained from our analysis can be beaten,
and our model will prove valuable as a component of a full, accurate model that incorporates realistic quantum memory in order to assess the reach of quantum-repeater-based QKD~\cite{DLCZ01} .

We develop our model of long-distance QKD by extending an existing 
practical protocol that modifies the BBM92 protocol using ES.
This practical protocol,
which is based on parametric down conversion (PDC) sources is given the acronym~PDC-ES-BBM92~\cite{SST11}.
Our modification here extends PDC-ES-BBM92,
which we call PDC-CES-BBM92,
to the case of multiple ES stations
by employing concatenated ES, or CES for short.
The earlier protocol PDC-ES-BBM92 is a special case of PDC-CES-BBM92
corresponding to having a single ES station. This special case corresponds to non-relay setup, with which we will compare our relay setups.

If a source were capable of producing a train of ideal Fourier-limited single photons~\cite{GVS04},
the BBM92 protocol would straightforwardly accommodate multiple repeater stations.
For practical QKD, which accounts for imperfect sources and detectors,
the protocol for sender~A and receiver~B must dictate actions for 
cases that higher counts are received
as described in the PDC-ES-BBM92 protocol.
The security of PDC-ES-BBM92 is assured through squashing,
with squashing referring to the effective projection of the protocol's Hilbert space to smaller dimension through the measurements conducted by~A and~B~\cite{TT08,BML08}.

For $N=3$ ES stations,
we show that our model predicts a secure communication distance 
of up to 850~km.
Unfortunately the rate becomes unimaginably low:
even at zero separation between~A and~B,
and simply taking into account just the device losses and the coincidence probability for the various PDC sources,
one bit would be generated every 150 billion years,
which is approximately ten times the presumed age of the universe.
Having~A and~B move far apart makes the rate even worse due to channel losses between PDC-based ES stations.
The pessimistic predictions of our model do not negate the value of having a realistic 
ES-based QKD description;
rather such a model is needed as a foundation for accurate descriptions of ultimately feasible long-distance QKD.

As our protocol performs badly with respect to SKGR,
we evaluate how well PDC-CES-BBM92 performs with respect
to the Takeoka-Guha-Wilde (TGW) bound for SKGR~\cite{TGW14}.
We Thus, determine the SKGR
considering ideal single-photon sources and unit-efficiency zero-dark-count detectors
thereby ensuring that our protocol is limited only
by channel loss and concatenations.
The resultant idealized SKGR for our protocol
is only slightly below the TGW bound,
which depends solely on channel loss regardless of optical power.
Thus, our protocol is reaching close to an bound for protocols that do not use quantum repeaters.

The paper is organized as follows.
In Sec.~\ref{sec:background}, we briefly explain the resources used in entanglement-based protocols.
We provide a brief account of BBM92 and PDC-ES-BBM92 protocols
and SKGR associated with these protocols,
and we discuss the SKGR TGW upper bound for non-repeater-based QKD protocols.
In Sec.~\ref{sec:evaluatingvisibility}, we describe in detail the protocol for evaluating visibility. 

In Sec.~\ref{sec:qberr},
we explicate how the quantum-bit-error rate (QBER) and the key rate are calculated.
We determine optimal parameter choices
for maximizing the SKGR,
and provide these optimal resource parameters in Sec.~\ref{sec:results}.
We present and plot our results for $N=2$ and $N=3$ stations for PDC-CES-BBM92 protocol. 
We compare our PDC-CES-BBM92 protocol with the decoy-state protocol,
and we find the SKGR upper bound for our protocol and compare it with TGW bound.
Finally, we conclude in Sec.~\ref{sec:conclusions}.

\section{Background}
\label{sec:background}
In this section, we provide a background on the imperfections of the optical resources
as well as describing existing entanglement-based protocols on which our PDC-CES-BBM92 protocol  is based.
Specifically these protocols comprise the BBM92 protocol and the PDC-ES-BBM92 protocol,
which are special cases of our new protocol. 
We also provide a brief account of the measurement-device-independent decoy-state protocol,
which extends furthest among the protocols studied Thus, far.
Finally we assess the TGW bound applicable for any QKD protocol not using quantum repeaters.

We consider PDC sources with photon-pair production rate~$\chi^2$.
The state of each PDC source can be represented by the pure state
\begin{equation}
\label{eq:PDC}
	\ket{\chi}
		=\exp\left[\text{i}\chi\left(\hat{a}^\dagger_\text{H}\hat{b}^\dagger_\text{H}
			+\hat{a}^\dagger_\text{V}\hat{b}^\dagger_\text{V}+\text{hc}\right)\right]	
\end{equation}
with~$\hat a$ and~$\hat b$ the annihilation operators
for the two spatial modes with two polarizations $\{H,V\}$,
and hc denoting Hermitian conjugate.

We consider threshold detectors with efficiency~$\eta$ and dark-count probability $\wp$. Highly-efficient detectors are desirable for long-distance QKD,
but, in practice, efficiency is traded against increasing dark counts.
For commonly used InGaAs detectors this trade off is
\begin{equation}
\label{eq:detector}
       \wp=A\text{e}^{B\eta},  
\end{equation}
with typical values~\cite{CGR05}
\begin{equation}
	A=6.1\times 10^{-7},\;
	B=17.
\end{equation}
A more efficient detector will,
in turn,
have a higher dark-count rate.

Ineficiency accounting for detector inefficiency~$\eta$ plus channel losses is
\begin{equation}
	10^{-(\alpha\ell+\alpha_0)/10}\eta
\end{equation}
for~$\ell$ the length of channel between the source,
and the detector.
The loss coefficient is~$\alpha$, 
which is given in units of dB/km,
and a fixed loss parameter~$\alpha_0$ incorporates all other distance-independent losses.

For the BBM92 protocol~\cite{BBM92} based on perfect resources,
the two legitimate users (sender~A and receiver~B)
share an entangled pair of qubits.
They measure the states of their own qubit in one of two bases,
chosen randomly.
Subsequently~A and~B reconcile a posteriori
their data strings by selecting those cases for which basis choices were identical.
Ultimately they obtain a shared set of random numbers from which they build a key,
and they use privacy amplification to ensure that messages can be sent securely from~A to~B.

The ES-BBM92 protocol modifies the BBM92 protocol by using entanglement swapping  to distribute entanglement between two users who have never interacted. Each of the two legitimate users have a source producing an entangled pair of photons in two modes.
A Bell state measurement (BSM) is performed on two modes, with one mode emanating from each of the two-mode sources.

A singlet state is post-selected after the BSM. The singlet state is just one of the four Bell states
\begin{equation}
\label{eq:bellstates}
	\ket{\Psi^\pm}:=\frac{\ket{00}\pm\ket{11}}{\sqrt{2}},\;
	\ket{\Phi^\pm}:=\frac{\ket{01}\pm\ket{10})}{\sqrt{2}}
\end{equation}
(with $\ket{0}$ and~$\ket{1}$ the logical qubit in the off and on states, respectively).
The ideal singlet measurement corresponds to the projector
$\ket{\Psi^-}\bra{\Psi^-}$. If the initial source is a perfect single-pair source, then the post-selection at the BSM projects the state of the remaining two modes into a singlet state. Since the two users now share entangled state, BBM92 protocol can be applied.

In the realistic scenario of PDC-ES-BBM92~\cite{SST11},
PDC sources are used and the detectors and channels have imperfections discussed above.
The extent of entanglement shared between the two users is quantified by measuring the visibility~$V$ of multi-photon coincidences between the two legitimate users.

As the measurement-device-independent decoy-state protocol~\cite{LCQ12, XQL+13} has Thus, far
the maximum reported achievable distance,
we compare the distance achieved by our PDC-CES-BBM92 protocol with that achieved  by the measurement-device-independent decoy-state protocol,
which is a variant of the earlier decoy-state protocol~\cite{MQZ+05}.
In the decoy-state protocol,
decoy pulses supplement signal pulses to detect adversarial attacks.
In the measurement-device-independent protocol,
signal pulses are sent by the two legitimate users to the untrusted adversary and the post selection is done on the basis of measurement results of the adversary.
The measurement results of the decoy pulses are used to estimate the error rate of the single photon pulses. 

TGW establish an upper bound on SKGR for non-repeater-based protocols~\cite{TGW14}.
Specifically TGW show that,
regardless of the optical power,
the SKGR upper bound,
\begin{align}
	R_\text{TGW}=\log_2\frac{1+10^{-\alpha\ell/10}}{1-10^{-\alpha\ell/10}},
\label{eq:TGW}
\end{align}
varies solely with channel length~$\ell$.
According to this bound,
any non-repeater-based QKD protocol with SKGR exceeding~$R_\text{TGW}$ is insecure.

\section{Evaluating visibility}
\label{sec:evaluatingvisibility}

The QKD protocol depends on sufficient visibility of multi-photon coincidences
between the two parties~A and~B separated by distance~$\ell$
who are connected by a quantum and a classical channel
as shown in Fig.~\ref{fig:setup}. Here $N=1$ corresponds to the single swap non-relay setup.
\begin{figure}
	\includegraphics[width=\linewidth]{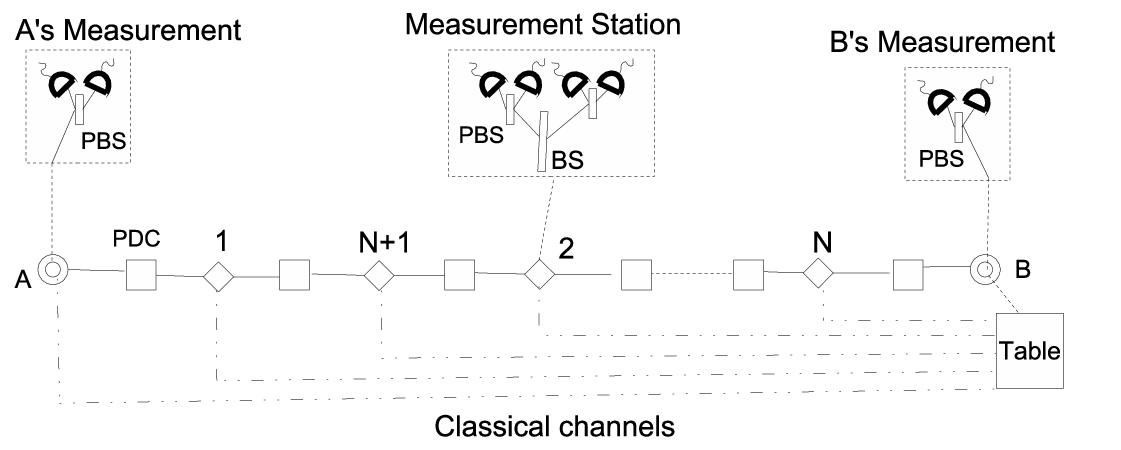}
\caption{
	Two parties~A and~B,
	separated by length~$\ell$,
	are connected by two channels: one quantum and the other classical.
	Alternating parametric down conversion (PDC) sources~$\square$
	and measurement stations~$\diamond$ periodically adjoin the channels.
	Two PDC sources and one measurement station comprise an entanglement swapping setup. For a total of $N$ swapping setups,  for~$x$  denoting the spatial co\"{o}rdinate from~A at $x=0$ to~B at $x=\ell$,
	measurement stations are located at each $x=\zeta\ell/(2N)$ for~$\zeta\in\{1,2,\dots,2N-1\}$,
	and PDC sources are located at each $x=(2\zeta+1)\ell/2N$ for~$\zeta\in\{0,1,2,\dots,N-1\}$.
	Insets:
	A and~B each have a polarizing beam splitters (PBS)
	to split the incoming beam into separate spatial modes for each polarization component,
	and a threshold detector for each of the two polarizations.
	B also has constructs a record of measurement events obtained by~A,
	by all measurement stations, and by himself.
	Each measurement station comprises a beam splitter (BS),
	two PBSs,
	and four threshold photon counters. 
	 }
\label{fig:setup}
\end{figure}
In this section we explain clearly how visibility should be evaluated for a CES scheme.
Although the concepts are known in the community,
our explanation in this section is valuable because each aspect of the protocol for visibility needs to be clear to establish the QKD protocol based on CES and to determine the protocol's security.

The co\"{o}rdinate extending from~A at the origin to~B is denoted~$x$.
To achieve long-distance QKD,
alternating, periodically-spaced PDC sources and measurement stations enable concatenated 
entanglement swapping so that~A and~B
can achieve shared entanglement despite photon losses en route.

The quantum channel is a medium, such as an optical fiber,
which is transport to light and allows superpositions of photon number and of polarizations to propagate 
non-negligible distances.
We treat the channel as having a spatially homogeneous loss rate,
and we neglect decoherence in the polarization basis.
The classical channel allows~A and measurement stations to transmit their measurement results to~B
so this channel can be oneway.

Each PDC source generates two entangled beams,
one propagating leftward (negative~$x$ direction)
and the other propagating rightward (positive~$x$ direction).
The light is picked up by a measurement station or by~A or by~B.
A and each measurement station transmit measurement results plus a time tag indicating the time of measurement through the classical channel to~B,
who uses these data to infer the photon-coincidence visibility~$V$.

Photodetection is achieved by employing threshold photon counters, 
which in the ideal case clicks only when at least one photon arrives and does not click
only when no photon is incident on the counter.
A detector click is sent to~B through the classical channel.
Due to detector inefficiency, an incident photon might not cause a click,
and, due to dark counts, the detector might click even if no signal photon has arrived.

As shown in an inset of Fig.~\ref{fig:setup},
each measurement station has a beam splitter (BS),
which accepts at its two input ports the leftward beam from the neighboring PDC source on the right
and the rightward beam from the neighboring PDC source on the left.
The two output beams are each directed to a polarizing beam splitter (PBS)
to separate the two polarizations into distinct spatial modes,
and these spatial modes are directed to four separate threshold photon counters:
two detectors on each side (left and right)
to register photons in the horizontal (H) and vertical (V) polarizations.

Also shown in the insets of Fig.~\ref{fig:setup},
A has a single incoming beam from the right, which she splits into two 
using a PBS and sends to two detectors.
B has the same set-up as A but additionally keeps a record of measurement data
transmitted to him by~A and each measurement station.

One or more detectors at a measurement station can click at any time.
When at least one detector clicks on each side (left and right),
the measurement station transmits to~B the time of the click and 
a binary signal indicating which detectors clicked.
As the measurement station hosts four detectors,
and each detector can click or not click at a given time,
eight possible detection events can occur at any time.
As at least one detector on each side (left and right) must click,
then the number of possible detection events reduces to seven possible recorded events.
The measurement station Thus, transmits to~B the time of the event and which possible outcome occurred. 

B collects all results and then creates time bins of short duration.
A detector click is stored as `1' and no click as `0',
and the bit is stored in the corresponding time bin.
The results from a station in one time bin Thus, form a four-bit string
with the first two bits allocated for~H and~V detectors on the left
and the last two for~H and~V detectors on the right.

The measurement results from~A and~B in one time bin are stored as a two-bit string from~H and~V detectors. 
B~post-selects those time bins in which measurement results from each station are anti correlated, implying a click in the~H detector on left and a click in~V on right and vice versa, which corresponds to string 1010 or 0110. 

For these post-selected time bins,
represented as lists,
B counts the coincidences in his and A's bit strings such that  A's string is 10 and his string is 01 or vice a versa. The number of all such coincidences is represented by the quantity `max',
as these events correspond to the maximum coincidences.
Minimum coincidences in the post-selected time bins occur when A's and B's strings are correlated.
This correlation implies detection event 10 or 01 at both occurred.
The number of all such time bins is recorded as `min'.
All the rest of the values in the time bins are discarded.

B calculates the visibility according to
\begin{equation}
\label{eq:V}
	V=\frac{\max-\min}{\max+\min}.
\end{equation}
Alternatively,
instead of discarding events~11 at any pair of detectors,
B could assign randomly either~01 or~10 to each~11 event.
The motivation for this procedure is to map higher-dimensional events,
due to multi-photon occurrences,
to a lower-dimensional space rather than just discard higher-dimensional events.
This assignment of higher-photon occurrences to lower-photon results is preferable for
security as this procedure enables the quashing map~\cite{MGB+10,GBN+14}.
A closed-form solution for~coincidence probability~\cite{KS14} is provided in Appendix~\ref{app:visibility}. 
 
\section{Quantum bit error rate and key-generation rate}
\label{sec:qberr}

The two figures of merit for QKD are QBER~$(Q)$ and SKGR ($R$).
We begin with a brief account of both as they relate to BBM92 and PDC-ES-BBM92 protocols. 
QBER~$Q$ is the ratio of wrong bits to the total number of bits
exchanged between the legitimate users,
and~$Q$ can be calculated from~$V$ according to~\cite{GRT+02}
\begin{equation}
\label{eq:QBERV}
	Q=\frac{1-V}{2}.
\end{equation}

The SKGR~$R$ is the ratio of bits retained between the legitimate users at the end of the QKD procedure to the bits sent originally
and serves as the figure of merit for achieving successful QKD.
This rate depends on two terms.
One is the sifted-key rate~$R_\text{sif}$ for events corresponding to~A and~B choosing the same bases.

The second term is the Shor-Preskill bound
\begin{equation}
\label{eq:shorpreskill}
        R_\text{SP}=1-\kappa H_2(Q)-H_2(Q)
\end{equation}
arising from optimal error correction and privacy amplification~\cite{SP00}
with
\begin{equation}
\label{eq:Shannonentropy}
       H_2(Q):=-Q\log_2Q-(1-Q)\log_2(1-Q)
\end{equation}
being the Shannon entropy.
The term $\kappa H_2$ is the lost-bit rate due to error correction,
with~$\kappa$ being the reconciliation efficiency and $\kappa=1$ for perfect reconciliation.
The SKGR is the product
\begin{equation}
\label{eq:keyrate}
       R=R_\text{sif} R_\text{SP},
\end{equation} 
which is approximately half the raw SKGR.

For PDC-ES-BBM92,
$R_\text{sif}$ depends on four factors:
\begin{enumerate}
	\item	each PDC source should successfully emit an entangled pair of photons;
	\item	the photons from each source should reach the BSM set up;
	\item	the BSM should be successful; and
	\item	the photons created at the two ends of the line must be detected by~A and~B.
 \end{enumerate}

For perfect reconciliation, $\kappa=1$.
The second term $H_2$ corresponds to key bits lost in privacy amplification. 
Above a certain cutoff (co) QBER denoted $Q_\text{co}(\kappa)$,
the rate bound~$R_\text{SP}$ and hence the rate~$R$ are no longer positive.
From Eq.~(\ref{eq:shorpreskill}), we can determine that
\begin{equation}
	Q_\text{co}(\kappa=1)=0.11,\;
	Q_\text{co}(\kappa=1.22)=0.094.
\end{equation}
We denote the maximum value of distance for which~$R$ is positive by $\ell_{\max}$.

The lower bound of $R_\text{SP}$~(\ref{eq:shorpreskill})
assumes perfect sources. However, Koashi and Preskill~\cite{KP03} proved that the same bound is valid for any defect in the source
as long as the defects do not reveal basis choice by legitimate users to the omnipotent eavesdropper.
This assumption holds even if a third party controls the source.

The rate bound is derived for qubits.
In a realistic setup, measurements are made on multi-photon states,
which span an infinite-dimensional Hilbert space.
The squashing techniques enables all measurements to be reduced to statistically equivalent measurements on qubits~\cite{BML08}.

Through squashing,
the incoming signal is first mapped (squashed) to a
two-dimensional Hilbert space whose physical state is a single-photon polarization.
The measurement is Thus, effectively performed on the single-qubit Hilbert space.
If the detection device is trusted and fully characterized,
then the squashing map can become part of the eavesdropper's attack. 
Therefore, the lower bound is the same as given by Eq.~(\ref{eq:shorpreskill}),
and all the deviations from perfect two photon source are 
accommodated by measuring the QBER~\cite{KP03}.

In order to calculate $R_\text{sif}$ in Eq.~(\ref{eq:keyrate}),
we calculate the four factors, which comprise raw SKGR for our set up:
\begin{enumerate}
	\item	Each PDC source successfully emits an entangled pair of photons with probability $(\chi^2)^{2N}$;
	\item	the photons from each source reach the BSM set up
with probability $10^{(-\alpha l/{40N})4N}$;
	\item	the BSM is successful with probability $\eta^2/2$
	bounded by its maximum value of $1/2$~\cite{CL01} so the net probability of $2N-1$
	successful BSM is $(\eta^2/2)^{2N-1}$; and
	\item	the photons at the two ends reach~A and~B with probability $\eta^2$.
\end{enumerate}
Thus, $R_\text{sif}$ is 
\begin{equation}
\label{eq:sifted}
     R_\text{sif}=\frac{1}{2}(\chi^2)^{2N}10^{(-\alpha l/{40N})4N}
     	(\eta^2/2)^{2N-1}\eta^2.
\end{equation}
The sifted-key rate decreases exponentially with increase in number of stations, $N$.~$R$ is then calculated from Eq.~(\ref{eq:keyrate}) using Eqs.~(\ref{eq:shorpreskill}) and~(\ref{eq:sifted}).

The SKGR for ideal resources, $R_\text{ideal}$,
is determined by treating perfect single-photon sources 
and perfect unit-efficiency zero-dark-count detectors
for which $V=1$.
Therefore, $R_\text{SP}=1$ so Eq.~(\ref{eq:keyrate}) shows that
the sifted-key rate,
$R_\text{sif}$
with $\eta=1$, is the sole contributor to the SKGR according to
\begin{align}
	R_\text{ideal}=R_\text{sif}=2^{-2N}10^{-\alpha\ell/10},
\label{eq:cesbound}
\end{align}
which must be less than the TGW bound~(\ref{eq:TGW}). 

\section{Maximizing the key rate}
\label{sec:results}
One goal of successful QKD is to maximize~$R$ for as large a distance as possible.
We have seen that,
in the case of the PDC-CES-BBM92 protocol,
$R$ depends on various resource parameters.

Most importantly~$R_\text{sif}$ and~$R_\text{SP}$ depend differently on~$\chi$; whereas~$R_\text{sif}$ decreases exponentially on decreasing~$\chi$, a low value of~$\chi$ is needed to maximize~$R_\text{SP}$. Both factors require low~$\wp$ and high~$\eta$ to be maximized.
We use numerical optimization technique to find the maximum value of~$R$,
$R_\text{max}$,
as a function of the resource parameters. 

In Subsec.~\ref{subsec:qberresources},
we analyze the dependence of QBER on PDC source efficiency~$\chi$.
Then, in Subsec.~\ref{subsec:optimalrate},
we obtain~$R_\text{max}$ over multiple resource parameters. We present and plot our numerical results for $N=2$ and $N=3$ and compare them with the known results of $N=1$. 
We also compare our protocol with the decoy-state protocol and compare
the idealized maximum key rate with the~TGW~bound.

\subsection{Dependence of QBER on resources}
\label{subsec:qberresources}
\begin{figure}
	\includegraphics[width=\linewidth]{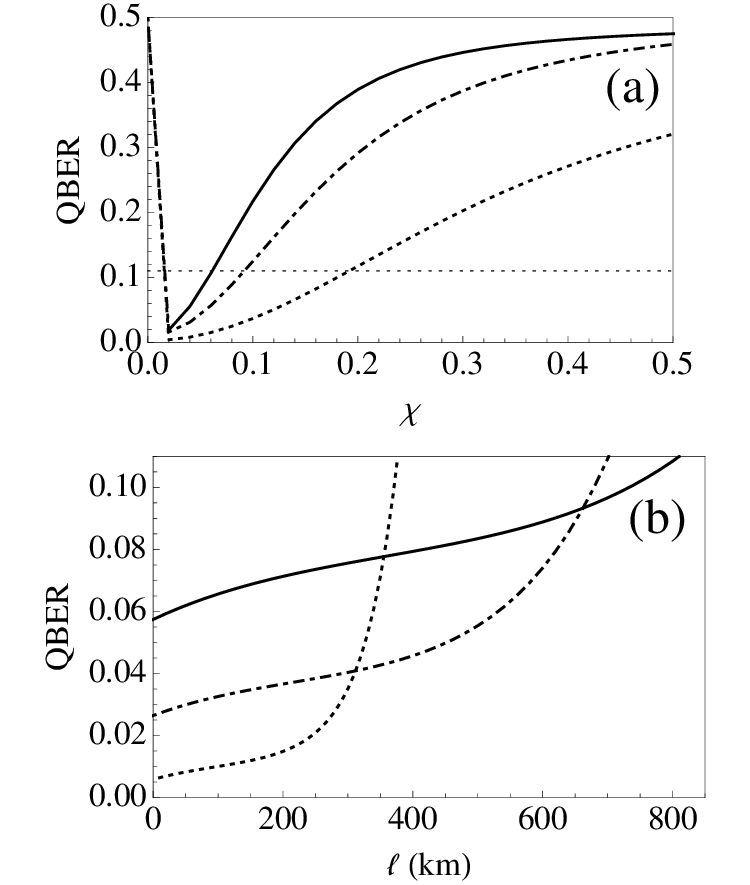}
\caption{
	QBER vs~$\chi$: QBER is plotted vs (a) source efficiency ($\chi$) and (b) distance ($\ell$) for $N=1$ (dotted curve), $N=2$ (dot-dashed curve) and $N=3$ (solid curve).
	Detector efficiency is set at $\eta=0.4$ and dark count efficiency is $\wp=1\times10^{-5}$.
     }
\label{fig:qbervschidist}
\end{figure}
QBER depends on~$V$ of the setup and hence in turn depends on all resource parameters.
We calculate~$V$~\cite{KS14} and analyze the dependence on source efficiency~$\chi$ in Fig.~\ref{fig:qbervschidist}(a) for one, two and three stations,
which are labeled $N=1$, 2 and 3 respectively.
In order to maintain the QBER below~$Q_\text{co}$, 
$\chi$ needs to be maintained below $0.20$ for $N=1$, below $0.08$ for $N=2$ and below $0.05$ for $N=3$.

In seeking high~$R$, QBER is not the only influencing factor.
Sifted SKGR~$R_\text{sif}$ increases
for higher values of~$\chi$ as shown in Eq.~(\ref{eq:sifted}).
Thus, a tradeoff is needed between the two rates.
To this end,
we find optimal~$\chi$, namely~$\chi_\text{opt}$,
in order to obtain~$R_\text{max}$ as described in Subsec.~\ref{subsec:optimalrate}.

Another point of interest is to be able to achieve as long distance as possible with low QBER.
We analyze how~$Q$ varies with distance for different numbers of concatenations.
Figure~\ref{fig:qbervschidist}(b) shows the variation with distance up to the threshold $Q_\text{co}$ of~$0.11$.
To obtain~$R_\text{max}$,
$Q$ must be constrained within the range shown for each selected distance~$\ell$.

\subsection{Maximum~$R$}
\label{subsec:optimalrate}
 We obtain~$R_\text{max}$
 for various resource parameters. These parameters include source brightness~$\chi^2$,
the detector intrinsic efficiency~$\eta$ and detector dark counts,
but we only maximize~$R$ over two parameters~$\chi$ and~$\eta$ as~$\wp$ is constrained (\ref{eq:detector}).
The two parts of~$R$, namely $R_\text{sif}$ and $R_\text{SP}$, depend in different ways upon~$\chi$
and~$\eta$.
Whereas maximizing~$R_\text{sif}$ requires keeping~$\chi$ high,
maximizing~$R_\text{SP}$ requires~$\chi$ to be low.
For low~$\chi$, dark counts become more pronounced;
hence detector efficiency should be kept small in order to have low dark count rate. 

To obtain~$R_\text{max}$,
we employ a multi-dimensional optimization technique,
specifically the quasi-Newton method,
over two parameters.
An appropriate initial point needs to be provided
so that the search for~$R_\text{max}$ 
commences with a positive value of~$R$ for each distance.
Prior information about~$Q$
obtained in Subsec.~\ref{subsec:qberresources} is important.

The maximized~$R$ obtained after optimizing over the two parameters~$\chi$ and~$\eta$,
i.e., determining $\chi_\text{opt}$ and~$\eta_\text{opt}$
is shown in Fig.~\ref{fig:optimalrate}(a)
\begin{figure}
	\includegraphics[width=\linewidth]{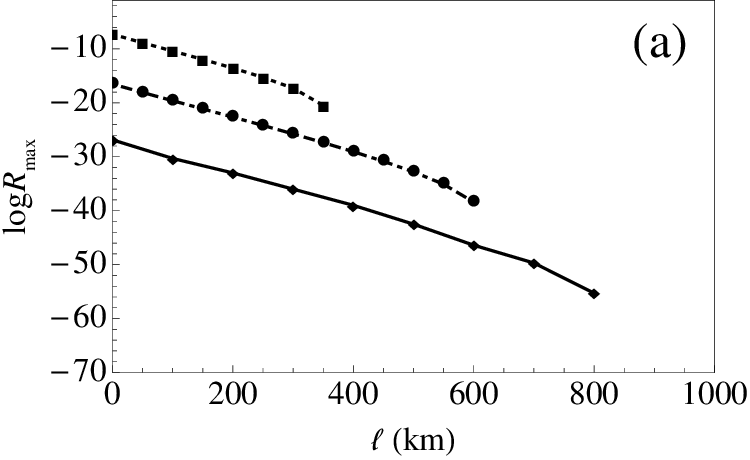}\\
	\includegraphics[width=\linewidth]{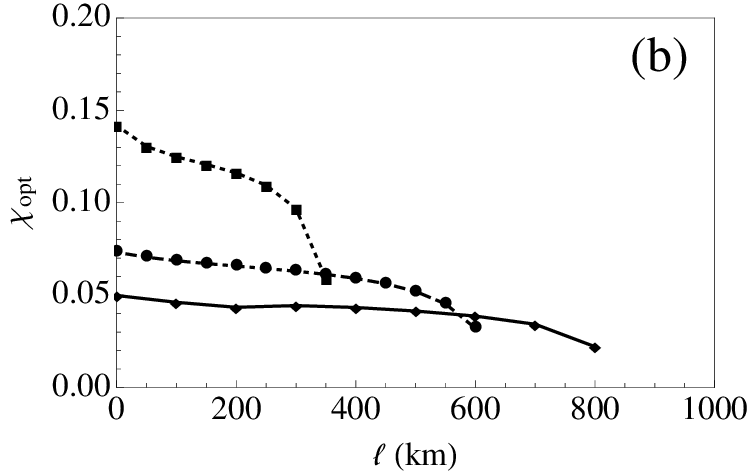}\\
	\includegraphics[width=\linewidth]{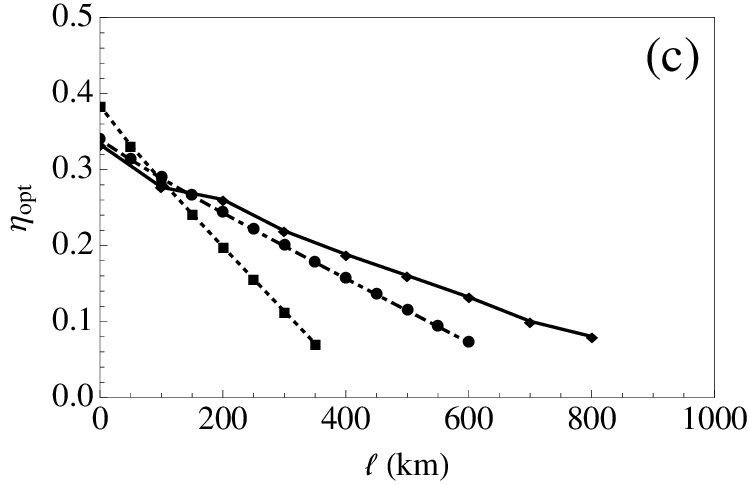}
\caption{
	Plot of (a)~$\log R_\text{max}$,
	(b)~$\chi_\text{opt}$
	and (c)~$\eta_\text{opt}$
	vs distance~$\ell$ up to maximum~$\ell_{\max}$
	for $N=1$ (dotted), $N=2$ (dot-dashed) and $N=3$ (solid)
	for $\alpha=0.25$~dB/km and $\alpha_0=4$~dB. 
	Symbols~$\blacksquare$,
	$\bullet$ and~$\blacklozenge$ represent numerically determined data points,
	which are connected by straight lines.
     }
\label{fig:optimalrate}
\end{figure}
for $N=1$, $N=2$ and $N=3$.
The loss coefficient is set at $\alpha=0.25$~dB/km, constant loss at $\alpha_0=4$~dB and $\kappa=1.22$~\cite{BS94}.
The optimization procedure requires calculating~$R$ many times, which is computationally expensive.

We observe a cutoff distance $\ell_{\max}\sim 350$~km for $N=1$.
For $N=2$ is $\ell_{\max}\sim 600$~km and,
for $N=3$,
$\ell_{\max}\sim 850$~km.
For larger distances, we failed to obtain a positive value of~$R$ despite
searching a wide swathe of parameters~$\chi$ and~$\eta$.
As can be seen from the plots,
$N=3$ should only be selected if $N=2$ and $N=1$ choices for concatenation fail.

The values for~$\chi_\text{opt}$ and~$\eta_\text{opt}$
that yield~$R_\text{max}$ in Fig.~\ref{fig:optimalrate}(a)
are presented in Figs.~\ref{fig:optimalrate}(b,c)
for the cases $N=1,2,3$.
As distance increases, $Q_\text{co}$ can become small if~$\chi$ is also small.
However, low values of~$\chi$ in turn require use of detectors with low efficiency in order to keep the dark count rate small.
This requirement is also clear from comparing the slopes of curves for~$\chi_\text{opt}$
and~$\eta_\text{opt}$.
The optimal~$\chi$ does not decrease for $N=2$ or for $N=3$ as rapidly as 
it does for $N=1$.

The maximum~$R$ obtained here is far too low for practical purposes.
Even for perfect detectors with unit efficiency and zero dark counts,
a low value of~$\chi$ is required to keep the QBER below $Q_\text{co}$.
This condition yields a maximum value of~$R$
corresponding to 1~bit each 10~msec for optimal $\chi_\text{opt}=0.2$
for a source with repetition rate 100~MHz at zero distance.
The corresponding rate for $N=2$ is one bit each 27.7~hrs for optimal 
choices~$\chi=0.1$.
For $N=3$ each bit requires waiting on average 
a mind boggling three million centuries for optimal choice $\chi=0.07$.
These sobering results emphasize the need to employ quantum memories~\cite{LST09, SAA+10}
or another protocol
to increase~$R$
to practical rates, and our theory serves as an important first step towards such accurate modeling.

In comparison with our protocol, the measurement-device-independent (MDI) decoy-state protocol~\cite{XQL+13} yields a communication distance of around 367~km with detector efficiency of 14\%.
Even for detector efficiency of 93\% and dark count probability of $10^{-6}$, the achievable distance is 667km with a PDC source in the middle. For fair comparison we find the maximized key generation rate for Superconducting Nanowire Single Photon Detectors (SNSPD) with high detector efficiency 93\% and low dark count rate~$10^{-9}$ \cite{MVS+13}, for our PDC-CES-BBM92 protocol with $N=2$ concatenations in Fig.~\ref{fig:snspd}. The achievable distance already reaches around~2000km, compared to around~800~km of MDI-QKD for the same parameters~\cite{PRXL14}.
Though the~MDI protocol yields higher key rates than our PDC-CES-BBM92 protocol and reaches much larger distance than BB84 protocol,
the achievable distance is much less than that achieved by PDC-CES-BBM92 protocol.
Eventually the low PDC-CES-BBM92 key rate could be mitigated
by incorporating quantum memory in a quantum-repeater setup.

In that case, which is the subject of future work,
such a quantum-repeater-based protocol could be superior to the decoy-state method for long-distance QKD.
\begin{figure}
	\includegraphics[width=\linewidth]{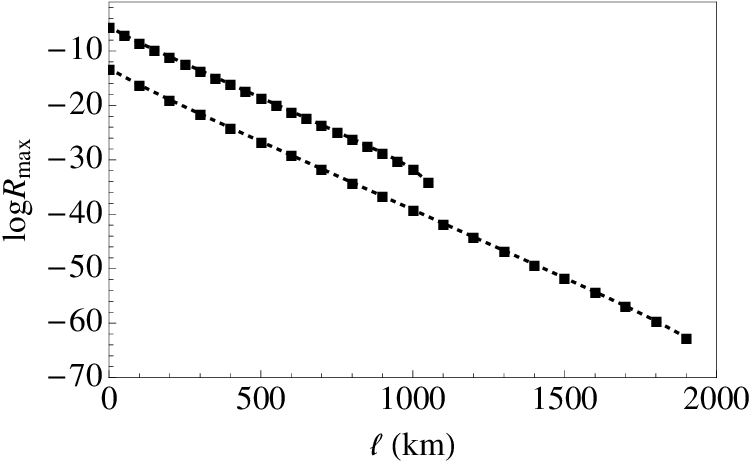}
\caption{
	Plot of $\log R_\text{max}$ for optimal $\chi$ for SNSPD with efficiency 93\% and dark count probability~$10^{-9}$ for $N=1$ (dotted) and $N=2$ (dot-dashed).
     }
\label{fig:snspd}
\end{figure}

\begin{figure}
	\includegraphics[width=\linewidth]{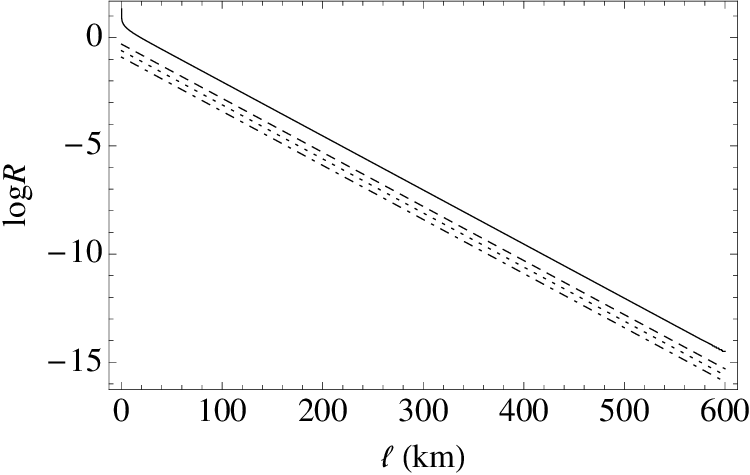}
\caption{
	Logarithmic plot of PDC-CES-BBM92 upper bound of~$R$ for $N=1$ (dashed curve), 
	$N=2$ (dotted curve),
	and $N=3$ (dot-dashed) for PDC-CES-BBM92
	compared with the TGW bound (solid curve)
	for various distances~$\ell$.
     }
\label{fig:boundcomp}
\end{figure}
The idealized key rate~$R_\text{ideal}$~(\ref{eq:cesbound}) for our PDC-CES-BBM92 protocol for $N=1,2,3$ for various distances are compared with the TGW bound~(\ref{eq:TGW}) in Fig.~\ref{fig:boundcomp}
with loss coefficient $\alpha=0.25$~db/km.
These idealized key rates for our protocol are below the TGW bound as expected; yet they are quite close to the TGW bound as well, which reiterates our earlier perception of necessity of including quantum memories for any further improvement in $R$. 
Multimode memories tend to compensate for low pair emission rate and hence lower key generation rate~\cite{KGD+15}

Although our theory is developed for arbitrarily many CES stations,
the exponentially increasing computational time with increasing number of concatenations
effectively limits our numerical analysis to $N\leq 3$ CES stations.

\section{Conclusions}
\label{sec:conclusions}

We have studied a long-distance quantum key distribution protocol 
that employs concatenated entanglement swapping~(CES) using parametric down conversion~(PDC) sources. Our model of this PDC-CES-BBM92 protocol accounts for realistic phenomena including multi-pair events,
inefficient detectors,
dark counts,
and lossy channels.
In particular, we have solved numerically 
for $N=2$ and $N=3$ concatenations
and compared these results with known results for the previously solved $N=1$ non-relay
entanglement-swapping setup.

Through our model and numerical methods,
we have determined optimal choices of resource parameters
including source efficiency,
detector dark-count rate,
and detector efficiency
so that these choices maximize the secret key-generation rate~$R$.
We have compared our results with the popular decoy-state protocol, which is a promising long-distance QKD protocol.
The key rate for perfect resources in our protocol are compared to the TGW bound, which is the upper bound for any QKD protocol not involving quantum memories. 

Our results show that secure key distribution distance can be increased upto~850~km using only three concatenated swapping setup,
which exceeds the achievable distance for the decoy-state protocol. 
We have found that for superconducting nanowire detectors the achievable distance is around 2000 km for $N=2$. Hence, low dark count rates and efficient detectors are desirable to deliver long-distance quantum communication.
However, because of the probabilistic nature of PDC sources, 
the key rates for the PDC-CES-BBM92 protocol are atrociously low.
On the other hand, for perfect resources these key rates approach closely to the TGW bound
so the low rate is close to the theoretical maximum for non-quantum-repeater-based protocols.
Therefore, quantum memories are needed to generate higher secret key-generation rates for probabilistic sources.
For fixed distance, visibility is always higher for fewer relays.
Our model provides a foundation for modeling future setups including all the resource imperfections,
and an important next step is to construct an accurate model for quantum memories and quantum repeaters.

\acknowledgments
We thank N.\ L\"{u}tkenhaus and E.\ Zahedinejad for valuable discussions and appreciate financial support
from the 1000 Talent Plan of China and from Alberta Innovates Technology Futures.
This research has been enabled by the use of computing resources
provided by WestGrid and Compute/Calcul Canada.

\appendix
\section{Coincidence probability for PDC-CES-BBM92 setup}
\label{app:visibility} 

As discussed in~Sec.~\ref{sec:evaluatingvisibility}, visibility is calculated from coincidence probability. Here we briefly explain the method to calculate coincidence probability at all $8N$ detectors. 

At each measurement setup in Fig.~\ref{fig:setup} there is a fourtuple of detectors, 
one for each~H mode and the other for each~V mode after the incident beam passes through each PBS.
For the sake of simplicity in the calculations,
we represent the four modes as $ijkl$, with~$i$ and~$j$ representing
the number of photons in the H and V modes, respectively,
on the left and~$k$ and~$l$ representing the number of photons in the V and H modes, respectively,
on the right of PBS in Fig~\ref{fig:setup}.

The conditional probability of clicks $ijkl$ on ideal detector such that actual detection event is $qrst$ is given by Baeysian approach~\cite{SHST09} as
\begin{align}
	P(\bm{ijkl}|\bm{qrst})=P(\bm{qrst|ijkl})P(\bm{ijkl})/P(\bm{qrst)},
\end{align}
where conditional probability of no-click at actual detector when perfect detector (with unit efficiency and zero dark counts) would have detected $i$ photons is given by 
\begin{equation}
	P(q=0|i)=(1-\wp)[1-10^{-(\alpha\ell+\alpha_0)/10}\eta(1-\wp)].
\end{equation}
The conditional probability of click is
\begin{equation}
	P(q=1|i)=1-P(q=0|i).
\end{equation}

The coincidence probability of clicks $q'r's't'$ at the detectors with~A and~B such that the detectors at the measurement stations give clicks $\bm{qrst}$
is 
\begin{align}
	P_\text{coinc}
		=&P(q'r's't|\bm{qrst})\nonumber\\
		=&\sum_{i'j'k'l',\bm{ijkl}}P(q'r's't'|i'j'k'l')
			\left|A^{\bm{ijkl}}_{i'j'k'l'}\right|^2
					\nonumber\\&\times
			P(\bm{ijkl}|\bm{qrst}),
\label{eq:coinc}
\end{align}
where $\bm q=\{q_1q_2\dots q_{2N-1}\}$ is a binary string with each $q=0$ representing a no-click and $q=1$ represents a click. Similar strings for~$\bm r$,~$\bm s$ and~$\bm t$ exist.

In Eq.~(\ref{eq:coinc}),
$\left|A^{\bm{ijkl}}_{i'j'k'l'}\right|^2$ is the probability that the perfect detectors with~A would have detected $i'$ and $j'$ photons and those at~B would have detected $k'$ and $l'$ after passing through the polarizer rotators conditioned on detection of $\bm{ijkl}$ photons at perfect detectors at measurement stations. Here $\bm i=(i_1i_2....i_{2N-1})$ for the $2N-1$ four-tuple of detectors, and similar strings pertain for $j$, $k$, and $l$. Thus, $A^{\bm{ijkl}}_{i'j'k'l'}$ is~\cite{KS14}
\begin{widetext}
\begin{align}
A^{\bm{ijkl}}_{i'j'k'l'}=&
    \left(\prod_{p=1}^{N}\frac{1}{\sqrt{2^{i_p+j_p+k_p+l_p}i_p!j_p!k_p!l_p!}}
    \frac{(\tanh\chi)^{i_p+j_p+k_p+l_p}}     
         {\cosh^{4N}\chi}
        \sum_{\mu_p=0}^{i_p}\sum_{\nu_p=0}^{j_p}\sum_{\kappa_p=0}^{k_p}
         \sum_{\lambda_p=0}^{l_p}
            (-1)^{\mu_p+\nu_p}{i_p\choose \mu_p}{j_p\choose \nu_p}{k_p\choose \kappa_p}{l_p
                \choose \lambda_p}\right)\nonumber\\
    &\times\prod_{n=1}^{N-1}
        \Omega(\mu_{n},\lambda_{n},i_{N+n},l_{N+n})
            \Omega(\nu_{n}\kappa_{n},j_{N+n},k_{N+n})
       \frac{\sqrt{i_{N+n}!j_{N+n}!k_{N+n}!l_{N+n}!}}       
        {(\sqrt{2})^{i_{N+n}+j_{N+n}+k_{N+n}+l_{N+n}}}\nonumber\\
    &\times\delta_{i_{N+n}+l_{N+n},    
        \mu_{n}+\lambda_{n}+i_{n+1}+l_{n+1}-
         \mu_{n+1}-\lambda_{n+1}}
         \delta_{j_{N+n}+k_{N+n},
          \nu_{n}+\kappa_{n}+j_{n+1}+k_{n+1}-
            \nu_{n+1}-\kappa_{n+1}}\nonumber\\
    &\times(\nu_N+\kappa_N)!(j_1+k_1-\nu_1-\kappa_1)!\sqrt{\frac{j'!k'!}{i'!l'!}}
         \sum_{n_a=0}^{\min[j',\nu_N+\kappa_N]}
            \sum_{n_d=0}^{\min[k',j_1+k_1-\nu_1-\kappa_1]}
             \left(\text{i}\tan\frac{\tilde{\alpha}}{2}\right)^{\nu_N+\kappa_N+j'-2n_a}\nonumber\\
    &\times\left(\cos\frac{\tilde{\alpha}}
           {2}\right)^{i'+j'-2n_a}\left(\text{i}\tan\frac{\tilde{\delta}}{2}\right)^{k'+j_1+k_1-\nu_1-\kappa_1-2n_d}
              \left(\cos\frac{\tilde{\delta}}{2}\right)^{l'+k'-2n_d}
          \nonumber\\
   &\times\frac{(i'+j'-n_a)!(l'+k'-nd)!}{n_a!n_d!(j'-n_a)!(k'-n_d)!(\nu_N+\kappa_N-n_a)! 
           (j_1+k_1-\nu_1-\kappa_1-n_d)!}\nonumber\\
    &\times\delta_{i'+j',
         \mu_N+\nu_N+\kappa_N+\lambda_N}
\delta_{k'+l',i_1+j_1+k_1+l_1-\mu_1-\nu_1-\kappa_1-\lambda_1},
\label{eq:A}
\end{align}
\end{widetext}
where~$\tilde\alpha$ and~$\tilde\beta$
are the angles of the polarizer rotators at~A's and~B's measurement setup, respectively.
We have used the expression
\begin{align}
\label{eq:Omega}
	\Omega(\mu_n,
            \lambda_n,i_{N+n},l_{N+n})
		=&\sum_{\gamma=0}^{\mu_n+\lambda_n}{\mu_n+\lambda_n\choose \gamma}
			\nonumber\\&\times
		{{i_{N+n}+l_{N+n}-\mu_n-\lambda_n }\choose {i_{N+n}-\gamma}}
			\nonumber\\&\times
		(-1)^{\mu_n+\lambda_n-\gamma},
\end{align}
and the form of~$\Omega(\nu_n,\kappa_n,j_{N+n},k_{N+n})$ is analogous.

The closed-form solution for $A^{\bm{ijkl}}_{i'j'k'l'}$~(\ref{eq:A})
provides the explicit expression for the probability of detecting the~$i'$ and~$j'$ photons 
at~A and 
the~$k'$ and~$l'$ photons at~B, such that $\bm{ijkl}$ photons are detected at all intermediate stations.
This expression is used in Eq.~(\ref{eq:coinc}) to calculate the maximum and minimum coincidences in Eq.~(\ref{eq:V}).

\bibliography{references}
\end{document}